\documentclass [5p, number] {elsarticle}
\usepackage{graphicx}
\usepackage{setspace}
\usepackage{lineno}
\usepackage{amsmath}
\usepackage{amsmath}

\title{Using pulse shape analysis to improve the position resolution of a resisitive anode microchannel plate detector}
\date{\today}

\author{Davinder Siwal}
\author{B.B. Wiggins}
\author{and R. T. deSouza \corref{cor1}}

\address{Department of Chemistry and Center for Exploration of Energy and Matter, \\ 
Indiana University, 2401 Milo B. Sampson Lane, Bloomington, Indiana 47408, USA}

\cortext[cor1]{desouza@indiana.edu}

\begin{document}
\begin{abstract}
Digital signal processing techniques were employed to investigate the 
joint use of charge division 
and risetime analyses for the resistive anode (RA) coupled to a microchannel plate detector (MCP). In contrast to the typical approach of using the
relative charge at each corner of 
the RA, this joint approach results in a significantly improved position resolution.
A conventional charge division analysis utilizing analog signal processing provides a position measured resolution of 170 $\mu$m (FWHM). 
By using the correlation between risetime and position we were able to obtain a 
measured resolution of 92 $\mu$m (FWHM),
corresponding to an intrinsic resolution of 64 $\mu$m (FMHM) for a single Z-stack 
MCP detector. 
\end{abstract}

\maketitle

\section{Introduction}
Position-sensitive  microchannel plate (MCP) detectors are a powerful and widely used tool in imaging of
electrons, photons and ions \cite{WizaNIM79}. Since their inception in the late 1950's \cite{Lampton81}, they have 
been used in a variety of applications ranging from ion-molecule reaction scattering experiments \cite{Wester14} to 
fast neutron radiography \cite{Tremsin11}. In this detector, 
an incident electron, photon, or ion ejects an electron from the micorchannel plate. 
The microchannel plate acts as
a continuous dynode electron multiplier with millions of independent channels providing a typical
amplification of $\approx$10$^3$. Two MCP plates (chevron configuration) or three 
plates (Z-stack configuration) can be stacked to achieve higher gains. 
Different methods exist for measuring the position of the electron cloud exiting the microchannel plate 
thus determining the position of the incident particle. The principal techniques are :
multi-anode \cite{Fraser84}, helical delay line \cite{Sobottka88}, induced signal \cite{deSouza12}, 
and resistive anode \cite{Lampton79}. 
The resistive anode (RA) technique is particularly appealing for its simplicity. The electron cloud emanating from
the MCP stack is incident on a two-dimensional resistive sheet. Readout of the charge at the 
four corners of the sheet provides a measure of position of the incident particle
{\it via} charge centroiding.
Utilizing this simple approach a position resolution of 134 $\mu$m (FWHM) for a chevron configuration 
has been achieved \cite{Wiza79}. 
Although for many application a 
chevron stack of two microchannel plates provides sufficient amplification, the detection of 
low-intensity signals near the one electron limit requires configurations involving three (Z-stack) or 
more MCPs. For a Z-stack configuration the best reported resolution is 
100-200 $\mu$m (FWHM) \cite{Downie93}. 
By use of more complex MCP stack configurations and through the use of retarding potentials,
position resolution as good as 50 $\mu$m (FWHM) has been realized \cite{Firmani82,Floryan89,Murakami10}. 
In this work, we investigate the use of pulse shape analysis to improve 
the position resolution obtained with a simple Z-stack MCP-RA detector.

\section{Experimental setup}

\begin{figure}
\begin{center}
\includegraphics[scale=0.5]{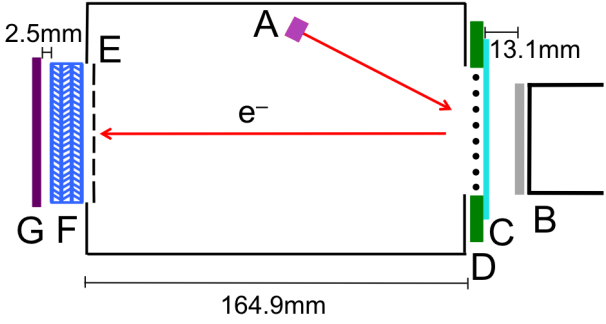}
\caption{
Experimental setup used to measure the position resolution of the Z-stack MCP-RA detector \cite{Wiggins15}. See text for details.}
\label{fig:setup}
\end{center}
\end{figure}

Depicted in Fig. \ref{fig:setup} is the experimental setup used to determine the position resolution of 
the MCP-RA detector. Alpha particles from a $^{241}$Am radioactive source ({\bf A}) are emitted towards a secondary electron emission
foil ({\bf C}). Electrons are emitted by the passage of an $\alpha$ particle through the 1.5 $\mu$m thick aluminized mylar foil. 
Ejected electrons
are accelerated to an energy of $\approx$ 1.3 keV by the potential difference between the aluminized foil and a wire grid ({\bf D}).
The electrons then pass through slits in a stainless steel mask ({\bf E}) directly in front of the microchannel plate stack ({\bf F}) 
before impinging on the front
surface of the MCP-RA detector. The precision mask, fabricated by laser micro-machining \cite{Potomac}, has ten slits 
each measuring 
100 $\mu$m by 7620 $\mu$m and spaced by either 4.2 or 4.5 mm apart.  
Electrons that pass through slits in the mask are amplified by the Z-stack MCP and detected on a resistive anode ({\bf G}). Alpha particles
traverse the aluminized mylar foil and are detected by a fast scintillator/photomultiplier tube (PMT) assembly ({\bf B}) placed 
directly behind the aluminized foil. 
The MCP used was a standard Z-stack (APD 3 40/12/10/12 D 60:1) with 10 $\mu$m diameter microchannels provided by 
Photonis USA \cite{PHOTONIS} which was coupled to a 
40 mm diameter RA from Quantar Technology Inc. \cite{QUANTAR}. 
Further details of the experimental setup can be found in \cite{Wiggins15}.

The entire assembly presented in Fig. \ref{fig:setup} is housed in a vacuum chamber that is evacuated to a pressure of 
4 x 10$^{-8}$ torr.
The microchannel plates were biased to a voltage of +3139 V using a ISEG NHQ224M low-noise, high voltage power supply (HVPS). 
The RA was biased to +3284 V, also using a ISEG NHQ224M HVPS. 
The secondary electron emission foil 
and photomultiplier tube were biased to voltages of -300 V and -1800 V using HK 5900 and Bertan 362 HVPS respectively.  
Signals detected at each corner of the resistive anode were amplified by a high quality charge sensitive amplifier (CSA) \cite{Davin01} operated
in vacuum. The four CSAs are situated approximately 13 cm from the MCP-RA to mimimize cable capacitance.
The output signals from the CSAs are coupled to a 250 MS/s digitizer (Caen DT5720B) \cite{CAEN}. 
Readout of the digitizer is triggered by the coincidence of a
fast signal extracted from the back of the MCP detector and the PMT. The digitizer is readout by a standard PC and waveforms are recorded for 
subsequent analysis. For the analysis subsequently described the waveforms 
associated with a total of 260,000 coincident triggers were recorded.

\begin{figure}
\begin{center}
\includegraphics[scale=0.35]{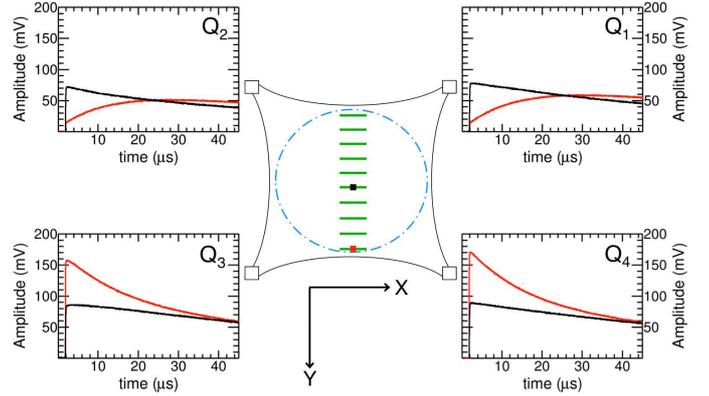}
\caption{
(Color Online) Schematic of the RA (black) along with the outline of the 40 mm diameter MCP (blue) and slits 
in the stainless steel mask (green). Pulse shapes associated with the electron
cloud incident on two locations of the RA are also shown. Slit 0 (s0) corresponds to the bottom slit and slit 9 (s9) to 
the topmost slit.}
\label{fig:RAPS}
\end{center}
\end{figure}

Indicated in Fig. \ref{fig:RAPS} is the reverse pincushion shape of the resistive anode (black) along with the circular outline of the 
40 mm diameter MCP detector (blue). Superimposed on the RA is the image of the ten slits provided by the stainless steel mask (green). 
The four corners of the 
MCP-RA are designated Q$_1$, Q$_2$, Q$_3$, and Q$_4$ as evident in Fig. \ref{fig:RAPS}. Along with the relative position of the MCP-RA and mask,
shown in Fig. \ref{fig:RAPS} are the signals measured at two locations on the RA. One set of digitized traces (red) correspond to
an electron cloud incident at
the the bottom slit of the RA as indicated by the red square. The other set of traces (black) correspond to a position close to the center
of the RA as indicated by the black square. The waveforms associated with these two positions are markedly different. When the position
signal arises from the center of the detector all the waveforms are essentially the same as is expected. However, when the signal 
originates at the bottom of the RA, while the two corners nearest the signal 
exhibit a fast risetime followed 
by an exponential decay, the upper corners of the RA manifest a significantly slower risetime. This risetime, dictated by the RC of the
resistive anode thus provides a measure of the particle's position. 
Motivated by prior work which utilized the risetime of signals in resistive silicon detectors to achieve position sensitivity in 
one dimension \cite{Kalbitzer67}, we elected to 
characterize the RA waveforms by their risetime.
The risetime (RT) of each signal was defined as the time required for the signal to go from 10$\%$ to 90$\%$ of its maximum value. As the slits
in the mask are oriented to probe the Y dimension of the RA, for the remainder of this work we focus on the position information 
in that dimension.

\section{Signal Risetime analysis}

Signals obtained from the CAEN digitizer are processed through a series 
of mathematical operations using a standard C++ code calling ROOT \cite{ROOT} 
libraries. In the present investigation the sampling resolution is 4 ns since  
the digitizer used has a sampling frequency of 250 MS/s. The obtained signals 
are corrected for the DC offset and gains for different channels for CSA's 
on an event-by-event basis.

\begin{figure}
\begin{center}
\includegraphics[scale=0.4]{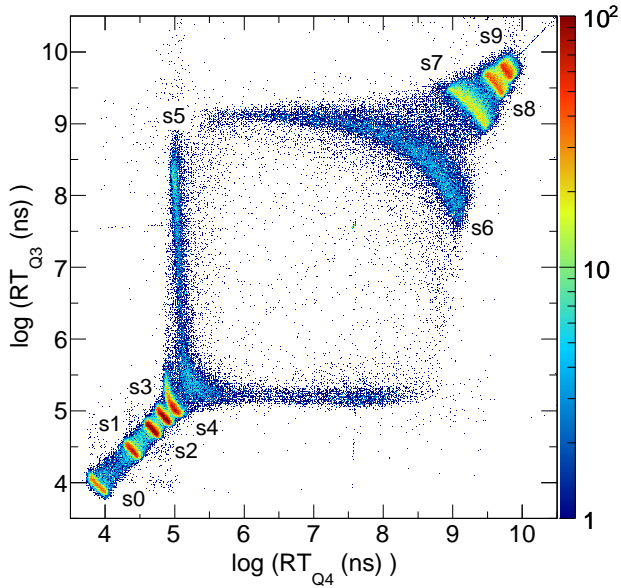}
\caption{
(Color online) Correlation between the risetime extracted from CSA output signals for the corners Q3 and Q4.
The numbering of the slits is indicated in the figure.}
\label{fig:RTcorr}
\end{center}
\end{figure}

Presented in Fig. \ref{fig:RTcorr} is the correlation between the risetime observed for signals at the two bottom corners if the RA namely
Q$_3$ and Q$_4$. Given the exponential dependence of the signal amplitude on the RC of the resistive anode, the correlation is
examined on a logarithmic scale. Individual slits are clearly evident in Fig. \ref{fig:RTcorr} with low numbered slits exhibiting shorter
risetimes and higher numbered slits associated with longer risetimes. An overall linear dependence between log(RT$_{Q3}$)
and log(RT$_{Q4}$) is observed as one moves from the bottom of the RA (s0) to the top (s9). Interestingly, a large jump in risetime is observed between
slit 5 (s5) and slit 6 (s6),
indicating a high sensitivity of the risetime to position in the center of the detector. 
Moreover, these two slits, in contrast to the other slits manifest a broad range of risetimes in at
least one of the two risetimes.  Electrons associated with these slits often do not show a strong correlation between the 
risetime of Q$_3$
and the risetime of Q$_4$ as indicated by the horizontal and vertical bands in the figure. 
Both the large sensitivity to position of the risetime in this region of the detector and the large dispersion in risetime
are likely due to the fact that the charge cloud incident in this region experiences near equal resistance in all directions and a relatively small
gradient in all directions.
For all but these center two slits however the 
risetime measured for the two corners exhibits a clear anti-correlation. This anti-correlation arises from the spatial extent of each slit in the x 
dimension. 
Motivated by the overall positive correlation between 
log(RT$_{Q3}$) and log(RT$_{Q4}$) we construct the quantity log(RT$_{Q3}$) + log(RT$_{Q4}$) corresponding to a line of unity slope
in Fig. \ref{fig:RTcorr}. It is clearly evident that the spacing of the slits along this line is not uniform.

\begin{figure}
\begin{center}
\includegraphics[scale=0.4]{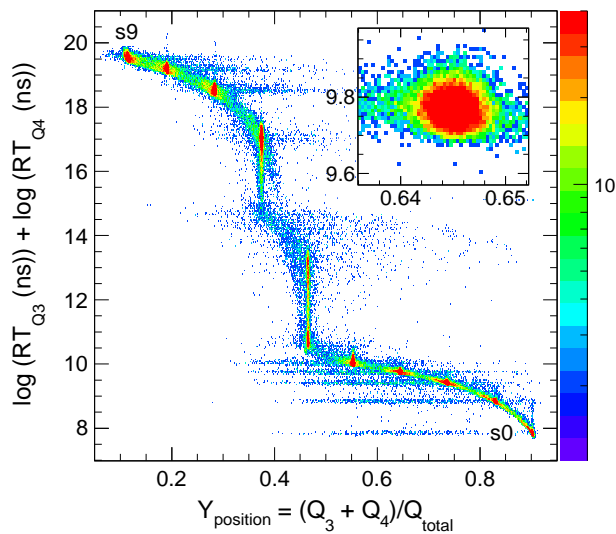}
\caption{
(Color online) Dependence of the summed risetime of Q3 and Q4 on the Y position of the electron 
cloud. The peaks corresponding to slit 0 (s0) and slit 9 (s9) are indicated. 
The peak associated with slit 3 is presented in the inset.}
\label{fig:RT_Ypos}
\end{center}
\end{figure}

In Fig.\ref{fig:RT_Ypos} the dependence of the summed risetime, 
log(RT$_{Q3}$) + log(RT$_{Q4}$), on the position in the Y dimension is explored. The position in the Y dimension is determined
by using the charge division method proposed by \cite{Fraser84,Lampton79}. 

\begin{equation}
Y_{position} = (Q_3 + Q_4)/Q_{total}
\end{equation}

\begin{equation}
Q_{total}=Q_1+Q_2+Q_3+Q_4
\end{equation}

To obtain the charge measured at each corner of the MCP-RA we have 
utilized digital filtering techniques to provide 
signal conditioning. We initially used a gaussian filter
on the digitized signal to integrate the signal from each CSA resulting 
in a gaussian-like pulse. The amplitude of this pulse is related to the
charge measured at the corner. To efficiently realize the gaussian filter 
a recursive algorithm was employed \cite{Smith97,Kihm03}. Use of an 
integration and differentiation time constant of 500 ns was determined 
to result in the best position resolution. The resolution obtained 
is comparable to the resolution of 170 $\mu$m obtained for this experimental 
setup with analog electronics \cite{Wiggins15}. 
To determine the sensitivity of our result to the filtering technique chosen 
we also used a trapezoidal filter. 
Further details on the signal processing can be found in the Appendix.

Evident in Fig. \ref{fig:RT_Ypos} is a clear correlation between the summed 
risetime and the $Y_{position}$ of the electron cloud. Since $Y_{position}$=0 is 
associated with the top of the RA, as the value of $Y_{position}$ increases 
one observes a general decrease in the summed risetime. This trend is 
understandable since as the position of the electron cloud moves closer to the
bottom of the RA the risetime decreases as initially evident in Fig. 
\ref{fig:setup}. The correlation observed for the upper half of the RA,
$Y_{position}$ $<$0.45 is also observed for the lower half of the RA,
$Y_{position} $$>$ 0.45. The fact that the observed locus is not centered on
$Y_{position}$=0.5 is likely due to the uncertainty in the 
positioning of the mask relative to the RA. Situated along 
the locus of points in Fig. \ref{fig:RT_Ypos}, that is 
the main feature of the spectrum, are a series of peaks corresponding to the 
slits in the mask. An enlarged region centered on slit 3 is displayed in 
the inset of the figure. The relationship between summed risetime
and $Y_{position}$ exhibited by the peaks in
\ref{fig:RT_Ypos} provides useful information. It describes how 
these two quantities should be related for a given position of the 
electron cloud on the RA. Selection of the data in these peaks corresponds
to approximately 25 $\%$ of the data. Aside from the strong locus with peaks visible in
Fig. \ref{fig:RT_Ypos} one also observes data that does not lie on this locus. 
The majority of this data is associated with horizontal lines which originate from
one of the ten peaks. Examination of the pulse shapes associated with this data 
clearly establishes that much of this data is associated with the pileup of two signals.
Such a pileup disturbs the baseline for the signal distorting the measured charge
while minimally disturbing the signal risetime.

\begin{figure}
\begin{center}
\includegraphics[scale=0.45]{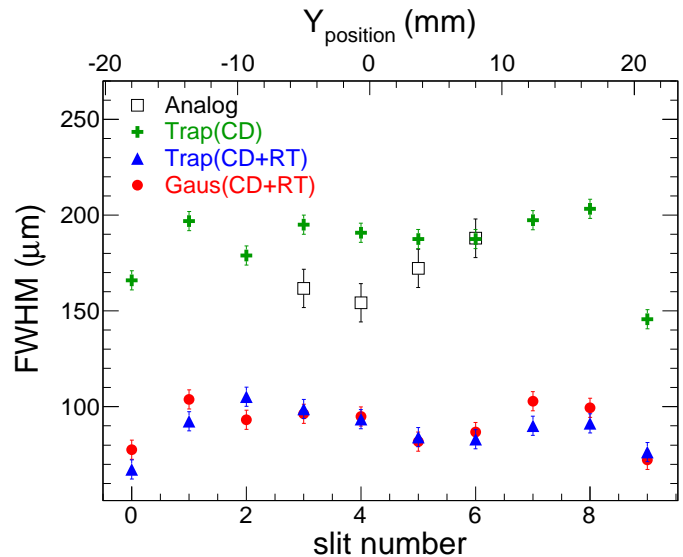}
\caption{
(Color online) Comparison of the measured resolution as a function of slit (position) using the 
correlation between risetime and position with the charge division 
approach alone. The resolution obtained using analog electronics is shown for 
reference.}
\label{fig:FWHM}
\end{center}
\end{figure}

To investigate if using the joint information from the summed risetime together
with the charge division method provides an improvement in the position
resolution we selected the ten peaks visible in 
Fig. \ref{fig:RT_Ypos}. As the resulting distributions are reasonably well described
by a gaussians, they were fit with this functional form. As the distance between the
slits corresponding to each of the peaks is well known the position spectrum was
calibrated and the position resolution extracted. The resulting resolution is 
presented in Fig. \ref{fig:FWHM}.

To begin we examine the measured resolution where only the charge division 
technique is utilized for the digitized signals. The result for this case 
which does not make use of any risetime information 
is presented as the 
cross symbol (green). The resolution obtained from this analysis with the 
trapezoidal filter is typically 
190 $\mu$m (FWHM). Use of the gaussian filter (not shown) gives a comparable result.
As the use of these filters mimics the use of an analog shaping amplifier, also 
shown for comparison is the resolution 
obtained from the use of analog electronics. Although the resolution obtained
with the analog electronics is $\approx$20 $\mu$m  lower than that 
obtained with the trapezoidal filter it is not a dramatic reduction. Moreover, 
using the analog electronics a reasonable resolution could only be obtained for the 
four central slits \cite{Wiggins15}. 
In contrast, using the digital signal processing approach a relatively uniform
resolution is obtained across the entire detector. The somewhat lower value 
observed for the two edge slits is due to the slight deviation of the 
distributions from gaussians.

The measured resolution obtained by the joint use of summed risetime along with 
charge division is depicted by the red circles (gaussian filter) and 
blue triangles (trapezoidal filter) respectively. Both
filters provide a measured resolution of $\approx$90 $\mu$m across entire 
detector. This result is a significant improvement over the use of the charge 
division approach alone. No systematic advantage is observed for one filter 
as compared to the other filter. Averaging the measured resolution of the central
eight slits for either the trapezoidal or gaussian filters, one obtains an average
measured resolution of 92 $\mu$m. Accounting for the finite slit width of 100 $\mu$m 
this measured resolution corresponds to an intrinsic resolution of 64 $\mu$m \cite{Wiggins15}.
This result is competitive with the resolution obtained using more complex MCP arrangements
and a retarding potential \cite{Firmani82,Floryan89,Murakami10}.

\section{Conclusions}
Resistive anode (RA) MCP detectors are widely used to provide position resolution in
the detection of electrons, photons, or ions. A charge division approach from the 
four corners of the RA is typically utilized to extract the centroid of the 
electron cloud emanating from the MCP. The signals arriving at the 
four corners of the RA manifest a broad range of risetimes. For a 40 mm diameter
MCP-RA detector, risetimes range from $<$ 1$\mu$s to approximately 40 $\mu$s.
By digitizing the signals from the MCP-RA with a high speed digitizer, 
digital signal processing techniques were utilized to extract both the 
charge collected at each corner as well as the risetime of the four signals 
for each incident electron. Implementation of the charge division 
approach resulted in a resolution of $\approx$190 $\mu$m (FWHM), 
comparable to that obtained with analog electronics.
Examination of the digitized signals  reveals a clear 
correlation between the risetime of the signals and the position of the electron
cloud. By using this correlation between risetime and position together with the 
charge division method a measured position resolution of $\approx$90 $\mu$m (FWHM) was achieved,
corresponding to an intrinsic resolution of 64 $\mu$m (FWHM). 
This result represents a significant improvement in the position resolution 
obtained. It was also observed that in the central region of the detector the 
risetime is extremely sensitive to the position of the electron cloud. While this 
latter observation requires further investigation for full characterization, 
it presents intriguing possibilities for enhanced imaging capabilities. 

A key question in image processing is the speed with which signals can be processed.
We therefore assessed the rate at which signals could be processed through our implementation of the 
digital filters described. Using a 2.8 GHz i686 PC we measured a rate of approximately 350 coincident events/s for
the trapezoidal filter and 200 events/sec for the gaussian filter. In the near term this processing 
rate could be increased by parallelization to make use of multiple cores on the computer. 
A longer term goal would involve implementation 
of FPGA processing at the detector level to both achieve a high processing rate 
and reduce the volume of data acquired.

\section{Acknowledgments}
We gratefully acknowledge the technical support provided by the personnel in the 
Mechanical Instrument Services and Electronic Instrument Services at the Department of Chemistry,
Indiana University.
This research is based upon work  supported by the U.S. Department of Energy, National Nuclear Security Administration 
under Award Number DE-NA0002012.

\section{Appendix: Signal Processing}

Standard digital signal processing techniques are used
on the CSA signals to extract the position resolution using
gaussian and trapezoidal filter algorithms \cite{Smith97,Jordanov94,Knoll94,Radeka72}. 
To efficiently process the pulse shapes,
recursive relations of both the filters were employed. For the gaussian shaper, the 
input signal is subjected to a combination of high-pass and low-pass filters. 
Each sample, $i$ of the $j^{th}$ channel CSA  
is subjected to a single-pole, high-pass digital filter  
using the
following recursive relation \cite{Smith97}\\

\begin{gather}
HP_{out}^{j}[i] = a_{HP}^{0}CSA^j[i]+a_{HP}^{1}CSA^j[i-1] \nonumber \\
+b_{HP}^{1}HP_{out}^j[i-1]
\label{HPEq}
\end{gather}
where $a_{HP}^0, \mbox{ }a_{HP}^1, \mbox{ and } b_{HP}^1$ are the coefficients 
of the high-pass filter kernel, which can be calculated from the following 
equations 

\begin{gather}
a_{HP}^{0} = (1+e^{-1/\Delta_{H}})/2 \\ 
a_{HP}^1 = -(1+e^{-1/\Delta_{H}})/2 \\
b_{HP}^1 = e^{-1/\Delta_{H}}
\end{gather}
$\Delta_{H}$ refers to the number of samples being considered for the 
high-pass time constant. The factor $e^{-1/\Delta_{H}}$ decides the 
sample-to-sample decay for a given time constant. Equation (\ref{HPEq}) mimics 
an analog differentiator circuit whose RC constant can be controlled by 
using the parameter $\Delta_{H}$ (in the units of number of samples). 
A time constant of 500 ns  was used for all the channels of CSA. Due to finite 
length of the exponential tail of input signal, the high pass filter output 
may overshoot the baseline of the pulse. This tendency can be 
corrected by using a pole-zero recursive 
relation \cite{Kihm03}. We utilized the
pole-zero recursive relation given by the following equation \\

\begin{equation}
HP_{corr}^{j}[i] =HP_{out}^{j}[i] + PZC \times CSA^{j}[i-1]
\end{equation}
where $PZC$ is a pole-zero correction factor. The value of $PZC$ 
can be chosen so as to correct for the overshoot. The 
second term provides the amplified (attenuated) exponential tail of the 
input pulse which corrects against the pole present in the exponential 
part of $HP_{out}^j$. The pole-zero corrected output of the high-pass filter 
is then transformed by a four-pole, low-pass digital filter, which is 
realized by the following recursive relations \cite{Smith97}

\begin{gather} 
GSH^{j}[i] = a_{LP}^{0}HP_{corr}^{j}[i]+b_{LP}^{1}GSH^{j}[i-1] \\ \nonumber 
+b_{LP}^{2}GSH^{j}[i-2]+ b_{LP}^{3}GSH^{j}[i-3] + b_{LP}^{4}GSH^{j}[i-4]
\end{gather}
where $a_{LP}^0, \mbox{ } b_{LP}^1, \mbox{ } b_{LP}^2, \mbox{ } b_{LP}^3, 
\mbox{ } b_{LP}^4 $ are the coefficients of a low-pass filter kernel 
and $GSH^{j}$ refers to the semi-gaussian shaper output obtained 
for the $j^{th}$ channel. 
The corresponding coefficients for the low-pass filter is given as

\begin{gather}
a_{LP}^{0} = (1-e^{-1/\Delta_{L}})^4 \\
b_{LP}^1 = 4e^{-1/\Delta_{L}} \\
b_{LP}^2 = -6(e^{-1/\Delta_{L}})^2 \\
b_{LP}^3 = 4(e^{-1/\Delta_{L}})^3 \\
b_{LP}=-(e^{-1/\Delta_{L}})^4
\end{gather}
where $\Delta_{L}$ is the low-pass time constant of the high-pass filter. 
For each corner of the resistive sheet, the measured charge is 
obtained by determining the 
height of the semi-gaussian shaper output.

To ensure that the extracted resolution was not limited by the particular filter
chosen, we also implemented a trapezoidal digital filter to extract the 
charge of the input signals. 
The following recursive relations are used for the $j^{th}$ channel CSA

\begin{gather}
d^{k,j}[i]=CSA^{j}[i]-CSA^j[i-k] \\
d^{k,l,j}[i] = d^{k,j}[i]-d^{k,j}[i-l]\\
p^j[i] = p^j[i-1] + m_2d^{k,l,j}[i]\\
r^j[i] = p^j[i] +m_1d^{k,l,j}[i]\\
s^j[i] = s^j[i-1] + r^j[i]
\end{gather}
Here $k$ represents the number of samples in the rising edge while $l$ is the 
total number of samples in the rising as well as in flat top region of the 
trapezoid shaped signal, $m_2$ is the gain of the filter, $m_1$ is the 
pole-zero correction factor, and $s^j$ is the output of the filter. 
The correction factor depends on the decay time constant ($\tau$) 
of the preamplifier, given by the following equation \cite{Knoll94}

\begin{equation}
m_1 = \frac{m_2}{e^{(T_{clk}/\tau)}-1}
\end{equation}
where $T_{clk}$ is the sample resolution of the digitizer. For shaping the 
CSA signals both the rising and flat top lengths are chosen to be 500 ns and the 
sample resolution is taken to be 4 ns.  The gain of the shaper is taken to 
be unity, while the decay time constant is 
chosen to be 30 $\mu s$ to match the decay constant of the CSA. Charge collected 
at each corner of the RA is calculated by taking the average height 
of the trapezoidal output.

\bibliographystyle{elsarticle-num-names}

\begin{thebibliography}{20}
\expandafter\ifx\csname natexlab\endcsname\relax\def\natexlab#1{#1}\fi
\providecommand{\url}[1]{\texttt{#1}}
\providecommand{\href}[2]{#2}
\providecommand{\path}[1]{#1}
\providecommand{\DOIprefix}{doi:}
\providecommand{\ArXivprefix}{arXiv:}
\providecommand{\URLprefix}{URL: }
\providecommand{\Pubmedprefix}{pmid:}
\providecommand{\doi}[1]{\href{http://dx.doi.org/#1}{\path{#1}}}
\providecommand{\Pubmed}[1]{\href{pmid:#1}{\path{#1}}}
\providecommand{\bibinfo}[2]{#2}
\ifx\xfnm\relax \def\xfnm[#1]{\unskip,\space#1}\fi
\bibitem[{Wiza(1979)Wiza}]{WizaNIM79}
\bibinfo{author}{J. L. Wiza}
\newblock \bibinfo{journal}{Nucl. Inst. and Meth.} \bibinfo{volume}{162}
  (\bibinfo{year}{1979}) \bibinfo{pages}{587}.
\bibitem[{Lampton(1981)Lampton}]{Lampton81}
\bibinfo{author}{M. Lampton}
\newblock \bibinfo{journal}{Sci. Am.} \bibinfo{volume}{245}
  (\bibinfo{year}{1981}) \bibinfo{pages}{62}.
\bibitem[{Wester(2014)Wester}]{Wester14}
\bibinfo{author}{R. Wester}
\newblock \bibinfo{journal}{Phys. Chem. Chem. Phys.} \bibinfo{volume}{16}
  (\bibinfo{year}{2014}) \bibinfo{pages}{396}.
\bibitem[{Tremsin et~al.(2011)Tremsin, McPhate, Vallerga, Siegmund, Feller, Lehmann, Butler,and Dawson}]{Tremsin11}
\bibinfo{author}{A. S. Tremsin et al.}, 
\newblock \bibinfo{journal}{Nucl. Inst. and Meth. A} \bibinfo{volume}{652}
  (\bibinfo{year}{2011}) \bibinfo{pages}{400}.
\bibitem[{Fraser(1984)Fraser}]{Fraser84}
\bibinfo{author}{G. W. Fraser},
\newblock \bibinfo{journal}{Nucl. Inst. and Meth. A} \bibinfo{volume}{221}
  (\bibinfo{year}{1984}) \bibinfo{pages}{115}.
\bibitem[{Sobottka and Williams.(1988)Sobottka and Williams}]{Sobottka88}
\bibinfo{author}{S. Sobottka}, \bibinfo{author}{M. Williams},
\newblock \bibinfo{journal}{IEEE Trans. Nucl. Sci.} \bibinfo{volume}{35}
  (\bibinfo{year}{1988}) \bibinfo{pages}{348}.
\bibitem[{deSouza et~al.(2012)deSouza, Gosser, and Hudan}]{deSouza12}
\bibinfo{author}{R. T. deSouza}, \bibinfo{author}{Z. Q. Gosser},
  \bibinfo{author}{S. Hudan},
\newblock \bibinfo{journal}{Rev. Sci. Instrum.} \bibinfo{volume}{83}
  (\bibinfo{year}{2012}) \bibinfo{pages}{053305}.
\bibitem[{Lampton and Carlson(1979)Lampton and Carlson}]{Lampton79}
\bibinfo{author}{M. Lampton}, \bibinfo{author}{C. W. Carlson},
\newblock \bibinfo{journal}{Rev. Sci. Instrum.} \bibinfo{volume}{50}
  (\bibinfo{year}{1979}) \bibinfo{pages}{1093}.
\bibitem[{Wiza et~al.(1979)Wiza, Henkel, and Roy}]{Wiza79}
\bibinfo{author}{J. L. Wiza}, \bibinfo{author}{P. R. Henkel},
  \bibinfo{author}{R. L. Roy},
\newblock \bibinfo{journal}{Rev. Sci. Instrum.} \bibinfo{volume}{48}
  (\bibinfo{year}{1979}) \bibinfo{pages}{1217}.
\bibitem[{Downie et~al.(1993)Downie, Litchfield, Parsons, Reynolds, and Powis}]{Downie93}
\bibinfo{author}{P. Downie et al.}, 
\newblock \bibinfo{journal}{Meas. Sci. Technol.} \bibinfo{volume}{4}
  (\bibinfo{year}{1993}) \bibinfo{pages}{1293}.
\bibitem[{Firmani et~al.(1982)Firmani, Ruiz, Carlson, Lampton, and Paresce}]{Firmani82}
\bibinfo{author}{C. Firmani}, \bibinfo{author}{E. Ruiz},
  \bibinfo{author}{C. W. Carlson}, \bibinfo{author}{M. Lampton},
  \bibinfo{author}{F. Parsece},
\newblock \bibinfo{journal}{Rev. Sci.Instrum.} \bibinfo{volume}{53}
  (\bibinfo{year}{1982}) \bibinfo{pages}{570}.
\bibitem[{Floryan and Johnson.(1989)Floryan and Johnson}]{Floryan89}
\bibinfo{author}{R. F. Floryan}, \bibinfo{author}{C. B. Johnson},
\newblock \bibinfo{journal}{Rev. Sci. Instrum.} \bibinfo{volume}{60}
  (\bibinfo{year}{1989}) \bibinfo{pages}{339}.
\bibitem[{Murakami et~al.(2010)Murakami, Yoshioka, and Yoshikawa}]{Murakami10}
\bibinfo{author}{G. Murakami}, \bibinfo{author}{K. Yoshioka},
  \bibinfo{author}{I. Yoshikawa},
\newblock \bibinfo{journal}{Appl. Opt.} \bibinfo{volume}{49}
  (\bibinfo{year}{2010}) \bibinfo{pages}{1293}.
\bibitem[{Mic(2015)}]{Potomac}
\bibinfo{title}{Potomac Photonics}, \bibinfo{year}{2015}. \URLprefix
  \url{http://www.potomac-laser.com}.
\bibitem[{Mic(2015)}]{PHOTONIS}
\bibinfo{title}{Photonis USA}, \bibinfo{year}{2015}. \URLprefix
  \url{http://www.photonis.com}.
\bibitem[{Mic(2015)}]{QUANTAR}
\bibinfo{title}{Quantar Technology Inc.}, \bibinfo{year}{2015}. \URLprefix
  \url{http://www.quantar.com}.
\bibitem[{Wiggins et~al.(2015)Wiggins, Richardson, Siwal, Hudan and deSouza}]{Wiggins15}
\bibinfo{author}{B. B. Wiggins et al.}, 
\newblock \bibinfo{journal}{Rev. Sci. Instrum.} \bibinfo{volume}{86}
(\bibinfo{year}{2015}) \bibinfo{pages}{083303}. 
\bibitem[{Davin et~al.(2001)Davin}]{Davin01}
\bibinfo{author}{B. Davin et al.},
\newblock \bibinfo{journal}{Nucl. Inst. and Meth. A} \bibinfo{volume}{473}
  (\bibinfo{year}{2001}) \bibinfo{pages}{302}.
\bibitem[{Mic(2015)}]{CAEN}
\bibinfo{title}{Caen Technologies, Inc.}, \bibinfo{year}{2015}. \URLprefix
  \url{http://www.caen.it}.
\bibitem[{Kalbitzer and Melzer(1967)Kalbitzer and Melzer}]{Kalbitzer67}
\bibinfo{author}{S. Kalbitzer}, \bibinfo{author}{W. Melzer},
\newblock \bibinfo{journal}{Nucl. Instr. and Meth.} \bibinfo{volume}{56}
  (\bibinfo{year}{1967}) \bibinfo{pages}{301}.
\bibitem[{R. Brun, and F. Rademakers (1997)}]{ROOT}
\bibinfo{author}{R.~Brun}, \bibinfo{author}{F.~Redemakers},
\newblock \bibinfo{journal}{Nucl. Instr. and Meth. A} \bibinfo{volume}{399}
  (\bibinfo{year}{1997}) \bibinfo{pages}{81}.
\bibitem[{S. W. Smith(1997)Smith}]{Smith97}
\bibinfo{author}{S.~W.~Smith},
\newblock \bibinfo{journal}{The Scientist and Engineer's Guide to Digital Signal Processing, California Technical Pub; 1st edition} 
ISBN 978-0966017632 \bibinfo{volume}{}
  (\bibinfo{year}{1997}) \bibinfo{pages}{319}.
\bibitem[{Kihm et~al.(2003)Kihm, Bobrakov, and Klapdor-Kleingrothaus}]{Kihm03}
\bibinfo{author}{T. Kihm}, \bibinfo{author}{V. F, Bobrakov},
  \bibinfo{author}{H. V. Klapdor-Kleingrothaus},
\newblock \bibinfo{journal}{Nucl. Inst. and Meth. A} \bibinfo{volume}{498}
  (\bibinfo{year}{2003}) \bibinfo{pages}{334}.
\bibitem[{V. T. Jordanov, G. F. Knoll (1994)}]{Jordanov94}
\bibinfo{author}{V. T. ~Jordanov}, \bibinfo{author}{G. F.~Knoll},
\newblock \bibinfo{journal}{Nucl. Instr. and Meth. A} \bibinfo{volume}{345}
  (\bibinfo{year}{1994}) \bibinfo{pages}{337}.
\bibitem[{V. T. Jordanov, G. F. Knoll et al.(1994)}]{Knoll94}
\bibinfo{author}{V. T. ~Jordanov},
\newblock \bibinfo{journal}{Nucl. Instr. and Meth. A} \bibinfo{volume}{353}
  (\bibinfo{year}{1994}) \bibinfo{pages}{261}.
\bibitem[{V. Radeka (1972)}]{Radeka72}
\bibinfo{author}{V. ~Radeka}, \newblock \bibinfo{journal}{Nucl. Instr. and Meth. A} \bibinfo{volume}{99}
  (\bibinfo{year}{1972}) \bibinfo{pages}{525}.

\end{thebibliography}

\end{document}